\documentclass{article}

\usepackage{arxiv}

\usepackage[utf8]{inputenc} 
\usepackage[T1]{fontenc}    
\usepackage{hyperref}       
\usepackage{url}            
\usepackage{booktabs}       
\usepackage{amsfonts}       
\usepackage{nicefrac}       
\usepackage{microtype}      
\usepackage{lipsum}		
\usepackage{graphicx}
\usepackage{natbib}
\usepackage{doi}

\title{Connectedness: A Dimension of Security Bug Severity Assessment For Measuring Uncertainty}

\date{March 14, 2025}	

\author{ {Chan Shue Long}\thanks{homepage: https://katsuragicsl.github.io/} \\
	\texttt{shue87655421@gmail.com} \\
}

\renewcommand{\shorttitle}{Connectedness}

\hypersetup{
pdftitle={Connectedness: A Dimension of Security Bug Severity Assessment For Measuring Uncertainty},
pdfsubject={cs.CR},
pdfauthor={Chan Shue Long},
pdfkeywords={Attack Surface, Cyber Risk Quantification, Philosophy of Cybersecurity, Uncertainty},
}

\begin{document}
\maketitle

\begin{abstract}
Current frameworks for evaluating security bug severity, such as the Common Vulnerability Scoring System (CVSS), prioritize the ratio of exploitability to impact. This paper suggests that the above approach measures the "known knowns" but inadequately addresses the "known unknowns" especially when there exist multiple possible exploit paths and side effects, which introduce significant uncertainty. This paper introduces the concept of connectedness, which measures how strongly a security bug is connected with different entities, thereby reflecting the uncertainty of impact and the exploit potential. This work highlights the critical but underappreciated role connectedness plays in severity assessments.
\end{abstract}


\section{Introduction}
\label{sec:intro}

The adage "we can not control what we can not measure." \citep{10.1145/1719030.1719036} underscores the centrality of quantification in cybersecurity. Yet, as Knight and Jones \citep{knight2002risk} observed, the conflation of risk and uncertainty persists \footnote{The practical difference between risk and uncertainty is "that in the former, the distribution of the outcome in a group of instances is known (either through calculation a priori or from statistics of past experience), while in the case of uncertainty this is not true" \citep{knight2002risk}. We can hereby define risk as the "known knowns" and uncertainty as the "unknown unknowns" as suggested by Simpson \citep{10.1093/cybsec/tyae022}. For example, a microservice having a 3 percent chance of melting down and being out of service is a risk; knowing that a feature of a given application is likely to be problematic without knowing its exact problems and how likely they occur, is an uncertainty.}. For instance, CVSS’s exploitability and impact metrics collapse multi-faceted vulnerabilities into singular values, disregarding the uncertainty inherent in unexplored attack vectors and side effects.

Today's severity assessment frameworks usually prioritize two dimensions: (1) the likelihood of a vulnerability being exploited and (2) the severity of its impact. The state-of-the-art scoring system CVSS measures them correspondingly with "exploitability metrics" and "impact metrics" \citep{mell2022measuring}\footnote{also see \url{https://www.first.org/cvss/v4.0/specification-document}}), and they are only capable of capturing the risk of a single exploit path. When there exist multiple possible exploit paths and multiple possible side effects, CVSS fails to address this situation and forces the user to use a single value (for each sub-item in the metrics) to represent the exploitability of multiple possible exploit paths and the impact of multiple possible side effects - in such cases, it fails to see a security bug as a set of possible security events and measure the effect of the induced uncertainty. In such situations, uncertainty plays an important role in severity analysis; hence, we need a way to measure the impact of uncertainty.

We will argue that connectedness is a good measure of uncertainty. First, we need to establish the view: "security bug is a set of possible security events".

\section{Security bug is a set of possible security events}
\label{sec:sec1}

A security bug can have multiple possible exploit paths and multiple possible side effects. Let us consider XSS as an example. In earlier days when XSS prevention techniques were not mature and feasible libraries were not abundant, people used custom filters to sanitize user input before reflecting it in the response, such as catching and removing dangerous elements in the input, HTML-encode the input, etc. However, these methods were not very effective, due to their bypasses of which many offensive security practitioners must be familiar with.

The main cause of the difficulties of the early XSS defense attempts faced is that there were many possible exploit paths, even if they were not known until someone published them: the user input reflected in the response body has connections with many different entities such as: 

\begin{enumerate}
    \item an HTML element
    \item an attribute of an existing element \footnote{such as the href attribute of an existing a tag}
    \item the inside of existing JavaScript code
    \item how HTML is parsed in different browsers and different libraries \footnote{for example see https://portswigger.net/research/bypassing-dompurify-again-with-mutation-xss}
\end{enumerate}

If we consider the impact of an XSS, we will also see multiple possible side effects, such as cookie stealing \footnote{which is usually the main concern.}, phishing \footnote{by disguising a phishing attempt as a genuine content served by a trusted domain, through its url.}, cross-origin cookie leakage \footnote{in case the cookie of the another site is scoped to parent domain, and the vulnerable site is another subdomain. This is hypothetical, but still a possible side effect, although not a strong one.}.

Hence, one should not see XSS as a single event, but as a set of possible events including:

\begin{enumerate}
    \item the user input creates a dangerous HTML entity
    \item the user input creates an entity with a dangerous attribute
    \item the user input creates a dangerous attribute in an existing entity
    \item the user input creates an entity which confuses the HTML parsing logic
\end{enumerate}

\section{Connectedness}
\label{sec:sec2}

\subsection{The more uncertainty, the more potential risk}
\label{sec:subsec1}

With hindsight, even if we lived in the early 2000s, we should have expected many possible ways to exploit XSS utilizing different objects related to the logic of how a webpage is rendered and how HTML is parsed. For example, even if we do not know that it is possible to exploit it by utilizing event handlers when the input is reflected in the \verb+href+ attribute of a \verb+<a>+ tag, we should still see it as a possible way. We do not know the actual way to exploit, but we do know that the behavior "user input getting reflected in the response" is connected with all components related to how a webpage is served, in particular the \verb+<a>+ tag and its \verb+href+ attribute; we just do not know exactly how and how likely it allows an exploit - this is the "known unknown". Once we find out an actual exploit, it becomes a "known known".

The more "known unknowns", the more uncertainty we have, and the more we should expect that some of them will one day turn into "known knowns". Equivalently, if a risk is a potential negative security event, we can see "uncertainty" as a potential risk.

\subsection{Definition of connectedness}

We hereby introduce the concept of \textbf{connectedness}: given a behavior (a system's behavior of interest: including security bugs), connectedness is a quantity describing the intensity of the behavior with other entities. The intensity consists of the number of entities connected with the given behavior and the strength of their connection. The more and stronger connections a given behavior has, the higher its connectedness. The strength of a connection is determined by how closely the given behavior is related to the entity.

In the case of XSS, the behavior of interest is that the user input pollutes the response and causes unexpected JavaScript execution. The entities connected with this behavior consist of HTML elements, attributes of existing elements, the inside of existing JavaScript code and the parser differentials \ref{sec:sec1}. The first three have strong connections with XSS as they would be directly affected by the user input - they are what would be served to the users; the last one has a weaker connection since its relation with XSS is more indirect.

To further illustrate the concept of connectedness, we study two examples in the next section.

\subsection{Examples}

\begin{enumerate}

    \item Input reflection

    Rated as informational usually (for example by Tenable\footnote{\url{https://www.tenable.com/plugins/was/114135}}). An input reflection can be seen as "a (reflected) XSS that did not make it". It connects to everything that a reflected XSS connects to, hence it has the same known unknowns, except that none of them has turned into an actual risk. However, as discussed in \ref{sec:subsec1}, there are multiple potential risks.

    \item Missing referrer policy

    Rated as informational usually (for example by Tenable\footnote{\url{https://www.tenable.com/plugins/was/98527}}). By default the policy will be \verb+strict-origin-when-cross-origin+ when the referrer policy header is not set. Comparing to the above issue, the possible exploit paths and the possible side effects are much lesser - obviously it depends on the threat. For example we assume that "leaking" the referrer to the same origin is not problematic; we also do not consider browser errors in managing the same origin policy (since in that case we could not do much about it as a webapp engineer; and much worse things would happen). If we assume that "leaking" the referrer to the same origin is a problem, for example path A and path B are served by different microservices, and for some reasons we do not want the microservices know which requests are directed from another microservice, then this referrer leak would be a problem. In usual settings, there are not many ways one can manipulate this issue. Hence its level of connectedness is comparatively lower than the above issue.\footnote{there are some possible side effects that were not mentioned, for example one could imagine that if there exists a bug in the server which causes DoS when a long referrer header is received, then the missing referrer policy might increasing an epsilon of chance of such DoS happens. But the connection between the missing referrer policy issue and this possible behavior is very weak (since the chance is small and such DoS could have been triggered by other means more efficiently), so it does not contribute much to the level of connectedness. I hope the reader is convinced that overall the input reflection issue still has higher level of connectedness.}

\end{enumerate}

Now we could see the difference made by adding connectedness into the analysis of severity: in the CVSS framework, both are usually treated as "informational" issues, inducing more or less the same level of risk.

However, they have very different levels of connectedness. If a security practitioner had to prioritize the triage/fix of one over that of the other, the input reflection issue should be prioritized for it has more potential risk. \footnote{The author also believes that the majority of security practitioners would do the same in such situation.}

By taking connectedness into account, one could differ security bugs with seemingly equal importance.

\subsection{Limitations}

The first limitation of applying connectedness on severity analysis is that whether an entity is considered as connected to the security bug in concern depends on the threat model of the system. If our threat model excluded a threat scenario, those entities that have meaningful interactions with the security bug only in that threat scenario will not be considered as connected with the security bug. Hence different threat models give different outcomes in the analysis of connectedness of the same security bug. See the footnote of example 2 in section 3.2.

The second limitation is that it is less objective to determine the strength of a connection, compared to current metrics, such as user interaction and privileges required, etc.

In defence of the first limitation, the author considers it a common problem of all severity analysis frameworks: security practitioners have to consider the threat model of the system in concern in order to accurately assess the severity of a given security bug. For example, if the system does not concern itself with the referrer being passed between different paths of the same origin, there is no point in considering its impact and possible exploits.

In defence of the second limitation, the author considers that in practice it is subjective enough for security practitioners to come to a consensus on the connectedness of a security bug, once they listed out and discussed all the connected entities. Indeed, it is possible for a security practitioner to miss certain entities and mis-evaluate the connectedness of a security bug, but this type of errors is in the category of "unknown unknowns"; also by discussions with other security practitioners they should realize what they missed. "Unknown unknowns" in severity analysis should not and could not be tackled with the methods of evaluation \footnote{since no matter what evaluation method one uses, by definition there will always be "unknown unknowns".}, but with something in a higher level of the abstraction ladder, such as continuous threat modeling, to (hopefully) reduce the "unknown unknowns".

\section{Conclusion}
\label{sec:secn}

In this paper, the author defined the concept of connectedness and showed how to apply it in the analysis of security bugs. The author also showed how it can differ security bugs that seemingly have the same severity in traditional frameworks.

There are limitations in the application of this concept: it depends on the threat model of the system. Also, while it helps us to evaluate the "known unknowns", it does not help us to tackle the "unknown unknowns": our analysis of connectedness on a given security bug could also be faulty - it is possible for us to miss significant exploit paths and side effects, and mistakenly believe that a particular security bug is of low connectedness.

However, it does not hurt the fact that connectedness helps assess the "known unknowns" and should be included in the severity analysis of security bugs. 

What we should take away from this paper is that uncertainty can strongly affect the severity of a security bug, and we should include connectedness as a measure of uncertainty, in order to achieve more accurate estimations of the severity of security bugs.

\section{Acknowledgement}

The author would like to thank Wong Wai Tuck for his comments on an early draft of this paper.

\bibliographystyle{unsrtnat}
\bibliography{references}

\end{document}